# Enhancing electrochemical intermediate solvation through electrolyte anion selection to increase nonaqueous Li-O$_2$ battery capacity

Colin M. Burke[1,2], Vikram Pande[3], Abhishek Khetan[4], Venkatasubramanian Viswanathan[3]* and Bryan D. McCloskey[1,2]*

[1]*Department of Chemical and Biomolecular Engineering, University of California, Berkeley, CA, 94720*

[2]*Environmental Energy Technologies Division, Lawrence Berkeley National Laboratory, Berkeley, CA, 94720*

[3]*Department of Mechanical Engineering, Carnegie Mellon University, Pittsburgh, PA, 15213*

[4]*Institute for Combustion Technology, RWTH, Aachen 52056, Germany*

*bmcclosk@berkeley.edu, venkvis@cmu.edu

**Significance Statement**
The Li-air battery has attracted significant interest as a potential high-energy alternative to Li-ion batteries. However, the battery discharge product, lithium peroxide, is both an electronic insulator and insoluble in nonaqueous electrolytes. It therefore passivates the battery cathode as it is uniformly deposited and disallows the battery to achieve even a modest fraction of its potential electrochemical capacity. Our research objective is to circumvent this challenge by enhancing the solubility of electrochemically formed intermediate species. We present a rational basis for electrolyte (i.e. solvent and salt) selection for nonaquoeus Li-air batteries and demonstrate a selection criteria for an electrolyte salt that increases the stability of Li+ in solution, thereby triggering a solution based process that allows significantly improved battery capacities.

## I. Abstract

Among the 'beyond Li-ion' battery chemistries, nonaqueous Li-O$_2$ batteries have the highest theoretical specific energy and as a result have attracted significant research attention over the past decade. A critical scientific challenge facing nonaqueous Li-O$_2$ batteries is the electronically insulating nature of the primary discharge product, lithium peroxide, which passivates the battery cathode as it is formed, leading to low ultimate cell capacities. Recently, strategies to enhance solubility to circumvent this issue have been reported, but rely upon electrolyte formulations that further decrease the overall electrochemical stability of the system, thereby deleteriously affecting battery rechargeability. In this study, we report that a significant enhancement (greater than four-fold) in Li-O$_2$ cell capacity is possible by appropriately selecting the salt anion in the electrolyte solution. Using [7]Li nuclear magnetic resonance and modeling, we confirm that this improvement is a result of enhanced Li+ stability in solution, which in turn induces solubility of the intermediate to Li$_2$O$_2$ formation. Using this strategy, the challenging task of identifying an electrolyte solvent that possesses the anti-correlated properties of high intermediate solubility and solvent stability is alleviated, potentially providing a pathway to develop an electrolyte that affords both high capacity and rechargeability. We believe the model and strategy presented here will be generally useful to enhance Coulombic efficiency in many electrochemical systems (e.g. Li-S batteries) where improving intermediate stability in solution could induce desired mechanisms of product formation.

## II. Introduction

The lithium-oxygen (Li-$O_2$) battery has garnered significant research interest in the past ten years due to its high theoretical specific energy (~3500 Wh/$kg_{Li2O2}$) compared to current state-of-the-art lithium-ion (Li-ion) batteries (~400 Wh/$kg_{LiCoO2}$).(1-4) Consisting of a lithium anode and an oxygen cathode, the nonaqueous Li-$O_2$ battery operates via the electrochemical formation and decomposition of lithium peroxide ($Li_2O_2$). The ideal overall reversible cell reaction is therefore:

$$2Li + O_2 \leftrightarrow Li_2O_2 \quad (2e^- \text{ process}) \ U = 2.96 \ V \quad 1).$$

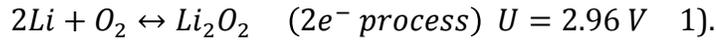

One challenge preventing the realization of the Li-$O_2$ battery's high theoretical specific energy is that the discharge product, $Li_2O_2$, which is generally insoluble in aprotic organic electrolytes, is an insulator.(5-8) As $Li_2O_2$ is conformally deposited on the cathode's carbon support during discharge, it electronically passivates the cathode, resulting in practical capacities much smaller than theoretically attainable.(9) Recently, two reports have described the engineering of electrolytes to circumvent this passivation and improve Li-$O_2$ battery discharge capacity. Aetukuri et al. suggests that adding ppm quantities of water to a DME-based electrolyte increases the solubility of intermediates during $Li_2O_2$ formation.(10) This increased solubility allows a reduced oxygen species shuttling mechanism that promotes deposition of $Li_2O_2$ aggregated toroid structures. The diffusion of the intermediates away from the electrode surface allows the surface to remain electronically accessible to $Li^+$ and $O_2$, promoting more $Li_2O_2$ growth, thereby leading to an increase in cell capacity. Of note, this increase in cell capacity with water content in the electrolyte is also consistent with reports by Gasteiger and coworkers.(11, 12) Aetukuri et al. reasons that this increase could be attributed to water's significantly higher Gutmann Acceptor Number (AN) than DME, as the AN is a measure of a solvent's Lewis acidity, and thus quantifies its ability to efficiently solubilize negatively charged species, such as the potential discharge product intermediate, superoxide ($O_2^-$).(13) Johnson et al. presents a related analysis, showing that an electrolyte solvent with a higher Gutmann Donor Number (DN), a measure of Lewis basicity,(14) is more likely to induce toroid formation due to increased $Li^+$ solvation efficiency, allowing solubility of $O_2^-$.(15)

While water and certain organic solvents increase cell capacity via this solution mechanism, there is evidence that both decrease electrolyte stability. Water impurities in Li-ion electrolytes are known to enhance parasitic electrochemical side reactions, and Aetukuri et al. and Cho et al. showed that adding ppm quantities of water in Li-$O_2$ batteries leads to a decrease in electrolyte stability and irreversible reactions with the lithium anode.(10, 16) Furthermore, using quantitative measures of battery rechargeability, high DN solvents, such as DMSO and N-methyl pyrrolidone, have been observed to be less stable than low DN solvents, such as acetonitrile and DME. (17) Recently, Khetan et al. showed using thermodynamic analysis that organic solvents' ability to induce the solution mechanism is anti-correlated with its stability towards nucleophilic attack.(18) Thus, Li-$O_2$ cells would benefit from an appropriately engineered electrolyte that both induces $Li_2O_2$ intermediate solubility and maintains or exceeds present electrolyte stability.

In this report, we describe the importance of the lithium salt anion in enhancing the solvation of electrochemically-formed intermediate species during Li-$O_2$ battery discharge, thereby enhancing discharge capacity. We present a study on two common Li-$O_2$ battery salts, lithium bis(trifluoromethane) sulfonimide (LiTFSI) and lithium nitrate (LiNO$_3$), dissolved in 1,2-dimethoxyethane (DME). These salts were selected because Schmeisser et al. found that TFSI$^-$ and NO$_3^-$ anions provided different DN in ionic liquids with common cations (NO$_3^-$-containing ILs having higher DN than TFSI$^-$-containing ILs). We also specifically selected NO$_3^-$ because of its reported positive influence on Li-$O_2$ battery rechargeability compared to the more commonly used TFSI$^-$.(19, 20) We found that electrolytes containing a high concentration of NO$_3^-$ exhibited higher donicity, as verified using $^7$Li NMR, and provided an increase in battery capacity greater than four-fold compared to a battery employing a TFSI$^-$ electrolyte, while not decreasing battery rechargeability, as measured using quantitative oxygen consumption and evolution. In order to theoretically quantify this enhancement, we developed an Ising model to describe the solvation shell of Li$^+$. This analysis shows the origin of this enhanced solution process is due to the formation of ion pairs (Li$^+$-NO3$^-$) in a DME solvent. The theoretical analysis further predicts that ion-pair formation and the associated enhancement in capacity will not be observed when DMSO is used as a solvent. We generalize this analysis to provide a rational basis for selection of electrolyte (solvent + salt) combinations for use in Li-$O_2$ batteries. We believe these results will have profound implications not only for Li-$O_2$ batteries, where a practical outcome of the solubility is an enhancement in battery capacity, but for other electrochemical systems (e.g. lithium-sulfur batteries) in which intermediate solvation may induce desired mechanisms of product formation.

## III. Results and discussion

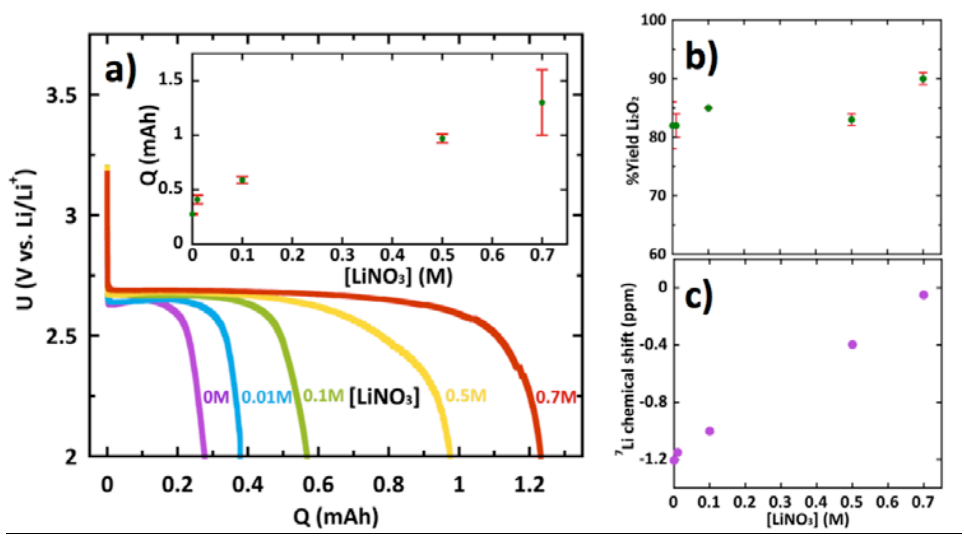

*Figure 1. a)* *Representative galvanostatic discharge profiles of Li-$O_2$ cells (450μA/cm$^2$ under a 1.5 atm $O_2$ atmosphere to a 2V cutoff). Inset shows capacity dependence on $LiNO_3$ concentration.* ***b)*** *$Li_2O_2$ discharge yield as a function of $LiNO_3$ concentration.* ***c)*** *$^7$Li chemical shift of sample electrolyte solutions, versus a 3 M LiCl in $D_2O$ standard, as a function of electrolyte $LiNO_3$ concentration. A less negative chemical shift represents a shift downfield. A 1.0M Li$^+$ concentration was used for all electrolytes (DME used as the solvent), while the LiTFSI/$LiNO_3$ ratio was varied. The $LiNO_3$ concentration for each cell is provided in the figure. As an example, the cell labeled '0.1M $LiNO_3$' contained 0.1M $LiNO_3$ and 0.9M LiTFSI. Error bars are one standard deviation of multiple experiments.*

To characterize the effects of the electrolyte salt anion on discharge performance, Li-$O_2$ cells were prepared with electrolytes of varying concentrations of $LiNO_3$ and LiTFSI salts, totaling 1.0M Li$^+$, in DME. Cell design and preparation is detailed in the Supporting Information and follows that described previously.(21)

Figure 1a presents representative galvanostatic discharge profiles of these Li-$O_2$ cells as a function of the $LiNO_3$ salt concentration. The inset in Figure 1a shows the average cell capacity for each LiNO₃ salt concentration. Cell capacity increases more than four-fold over the LiNO₃ concentration range studied, clearly indicating the substantial effect of the Li⁺ counterion on cell capacity.

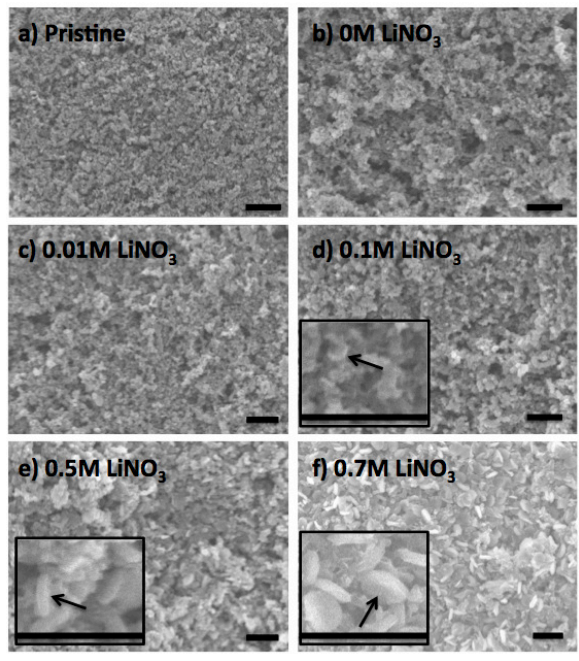

*Figure 2.* (a) Scanning electron microscope (SEM) image of pristine XC72 carbon cathode prior to discharge. (b-f) Discharged cathodes in cells employing 1M total Li⁺ concentration, with 0M LiNO₃ (1M LiTFSI), 0.01M LiNO₃, 0.1M LiNO₃, 0.5M LiNO₃, 0.7M LiNO₃, respectively. Cells were discharged at 45µA/cm² to 0.9mAh/cm² or a 2V cutoff voltage, but all cells had at least 0.5 mAh/cm² capacity. Scale bars are 1µm.

Scanning electron microscopy (SEM) was performed on discharged cathodes to investigate changes in $Li_2O_2$ morphology, and hence changes in discharge mechanism, with increased LiNO₃ concentration. Figure 2 presents SEM images of a pristine cathode (Figure 2a) and images of cathodes from cells of identical electrolyte compositions as those studied in Figure 1, but discharged at 45 µA/cm² (Figure 2b-f). When comparing Figure 2a-c, the pristine, 0M LiNO₃, and 0.01M LiNO₃ cathodes appear indistinguishable. This implies a conformal coating of discharge product on the 0M LiNO$_3$ and 0.01M LiNO$_3$ cathodes, and is consistent with previous reports for 1M LiTFSI in DME.(10, 22)

A conformal coating of discharge product is indicative of a thin film Li$_2$O$_2$ surface deposition mechanism. Originally outlined by Laoire et al., this mechanism is described by the following elementary steps (23, 24):

$$Li^+ + e^- + O_2^* \leftrightarrow LiO_2^* \quad (2)$$

$$Li^+ + e^- + LiO_2^* \leftrightarrow Li_2O_2^* \quad (3a)$$

and/or

$$2LiO_2^* \leftrightarrow Li_2O_2^* + O_2 \quad (3b)$$

where '*' denotes a species adsorbed to the cathode/Li$_2$O$_2$ surface. Importantly, in the DME/LiTFSI electrolyte, LiO$_2$* is insoluble and therefore remains adsorbed to the electrode surface, where a second charge transfer step (reaction 3a) or a disproportionation reaction (reaction 3b) results in the conformal Li$_2$O$_2$ coating observed in Figure 2b and c. (3, 22-26)

As the LiNO$_3$/LiTFSI ratio increases, the discharge morphology perceptibly changes. As seen in Figure 2d, when using 0.1M LiNO$_3$, nodular morphologies appear on the cathode surface. Increasing the LiNO$_3$ concentration to 0.5M and 0.7M finds these structures replaced with increasingly larger toroid structures, as seen in Figure 2e and f, respectively.

As described in the literature, the toroid morphology observed in Figure 2d-f is indicative of a solution mechanism of Li$_2$O$_2$ growth proceeding through solubility of the LiO$_2$ intermediate.(10, 15, 28) The solvation of LiO$_2$ into lithium cations and the redox active superoxide anion, O$_2^-$, follows the equilibrium reaction (10):

$$LiO_2^* \leftrightarrow Li^+(sol) + O_2^-(sol) \quad (4)$$

Solvated $O_2^-$ can then diffuse in solution to a growing $Li_2O_2$ toroid, where it combines with $Li^+$ from the solution to form adsorbed $LiO_2^*$ on the toroid surface. $LiO_2^*$ subsequently undergoes disproportionation according to Eq 3b, leading to the formation of $Li_2O_2$ on the toroid surface.(10) The observed toroid formation on discharged cathodes from cells employing high $LiNO_3/LiTFSI$ ratios supports the enhancement of this solution mechanism with increasing $LiNO_3$ concentration.

In further support of the solution mechanism, increasingly larger toroid structures were observed as current density decreased in cells employing 0.5M $LiNO_3$ (0.5M LiTFSI) (Figure S1). This observation is consistent with previous reports where $Li_2O_2$ toroid formation was correlated to the enhancement of the solution $Li_2O_2$ formation mechanism at low currents.(10, 15, 28)

Of note, we find that $Li_2O_2$ yield, as measured using an established peroxide titration technique,(27) is generally unaffected by the electrolyte compositions studied here (Figure 1b), although a slightly higher $Li_2O_2$ yield may be observed at high $LiNO_3$ concentrations. Differential electrochemical mass spectrometry (DEMS) was also used, as described previously,(21, 27) to quantify the reversibility of the electrochemical reactions (Figure S2). The ratio (OER/ORR) of the amount of oxygen evolved during charge (OER) to the amount of oxygen consumed during discharge (ORR), an important metric of reversibility, is statistically equal for a cell employing 1M LiTFSI and a cell employing 0.5M $LiNO_3$/0.5M LiTFSI (OER/ORR ~ 0.83). Furthermore, only $^{18}O_2$ is evolved on charge after a discharge under $^{18}O_2$ of a cell employing 0.5M $LiN^{16}O_3$/0.5M LiTFSI, confirming that $NO_3^-$ does not participate in the electrochemical reaction other than to induce solubility of the intermediates. This result agrees with a similar experiment using pure LiTFSI-based electrolytes,(22) implying that electroactive $O_2$ remains associated during both $Li_2O_2$ formation and oxidation.

With a change in anion clearly inducing a solution $Li_2O_2$ growth mechanism, it can be reasoned that the anion, a strong electron donor, is affecting $LiO_2$ solubility via enhanced $Li^+$ solvation. The electrolyte anion can affect the overall electrolyte's donicity (quantified by the Gutmann Donor number (DN), a measure of Lewis basicity(14)), in turn affecting its ability to solubilize $LiO_2$ through enhanced solvation of $Li^+$. We used $^7Li$ NMR to probe the electron donicity felt by $Li^+$ ions in our $LiNO_3$/LiTFSI in DME electrolytes as a proxy measurement of the relative effect of the anion on electrolyte DN.

Using NMR as a proxy for DN is a well-known technique, with Erlich et al. first proposing $^{23}Na$ NMR as an effective measurement for a solvent's DN. (29) Erlich et al. reasoned that a downfield $^{23}Na$ shift resulted from stronger interaction between the solvation shell molecules and the cation, thereby decreasing the cation's shielding. The environment of $Li^+$ in $LiNO_3$/LiTFSI in DME electrolytes cannot be determined via $^{23}Na$ NMR, though, as adding $NaClO_4$ to the electrolytes causes a white precipitate to crash out of solution (likely $NaNO_3$, as dissolving $NaClO_4$ in an anhydrous solvent containing $LiNO_3$ has been proposed as a method for making anhydrous $LiClO_4$ (30)). However, we reason that $^7Li$ NMR, in place of $^{23}Na$ NMR, can serve as a reasonable proxy of the relative donicity of $Li^+$ electrolytes in a single solvent.

Figure 1c shows the $^7$Li chemical shift, referenced to an external standard of LiCl in D$_2$O, of each LiNO$_3$/LiTFSI in DME electrolyte. As LiNO$_3$ concentration increases, the $^7$Li peak shifts downfield, or becomes less shielded. Cahen et al. showed that the $^7$Li chemical shift of a lithium salt may display a concentration dependence, contingent, to a first approximation, on the DN of the solvent and the DN of the anion.(31) The DN of an electrolyte containing a low DN solvent and high DN anion, like Br$^-$ (DN=33.7) in acetonitrile (DN=14.1), exhibits an anion concentration dependence (DN values from Linert et al.(32)). Conversely, electrolytes comprised of a high DN solvent with a relatively low DN anion, like ClO$_4^-$ (DN=8.44) in dimethyl sulfoxide (DN=29.8), do not exhibit a DN dependence on anion concentration. These trends agree with Linert et al., who found via solvatochromic dyes that the effective DN of an electrolyte depended on an interplay between the DN of the solvent, DN of the anion, and AN of the solvent.(32) For example, if the solvent's DN was larger than the anion's DN, then the electrolyte comprising the two had a DN similar to its solvent's DN.

It can be reasoned here that if LiNO$_3$ indeed has a higher DN than DME, then increasing the concentration of LiNO$_3$ will increase the number of NO$_3^-$ interacting with any particular Li$^+$, which, in turn, will lead to an increase in the electrolyte's DN. Thus, we reason that the presence of a concentration dependence on $^7$Li chemical shift indicates NO$_3^-$ serving an active role in the electrolyte's donicity, and the increasingly downfield shift of $^7$Li with increasing LiNO$_3$ concentration represents increasing donicity. In contrast, Figure 3 shows that indeed LiNO$_3$/LiTFSI salts in the high DN solvent dimethyl sulfoxide do not exhibit a substantial change in

$^7$Li shift with increasing LiNO$_3$ concentration, and, as therefore expected, no statistically significant capacity increase is observed in DMSO-based electrolytes as the LiNO$_3$/TFSI ratio increases.

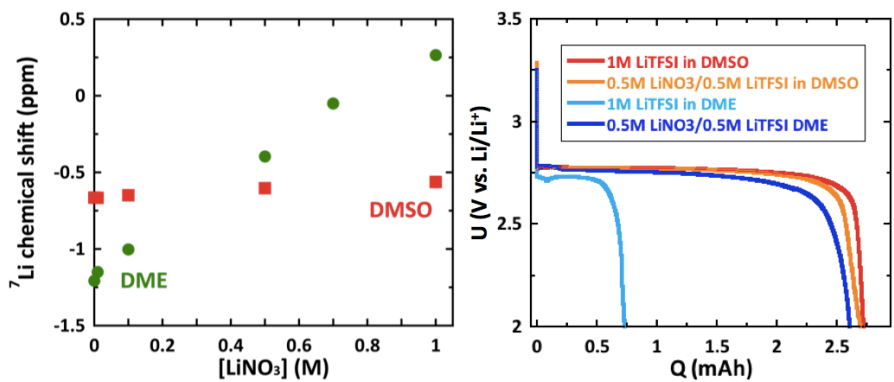

**Figure 3. a)** $^7$Li chemical shift of DMSO and DME-based electrolytes, versus a 3 M LiCl in D$_2$O reference, as a function of electrolyte LiNO$_3$ concentration. A less negative chemical shift represents a shift downfield. **b)** Discharge profiles (45 µA/cm$^2$, 2V cutoff) as a function of LiNO$_3$ concentration for DMSO and DME-based electrolytes. A 1.0M Li$^+$ concentration was used for all cell electrolytes, while the LiTFSI/LiNO$_3$ ratio was varied. The LiNO$_3$ concentration for each cell is provided.

We note, however, that extreme care must be taken when using NMR techniques to compare and quantify solvent DNs, particularly between dislike solvent classes, such as protic and aprotic solvents, as was discussed by Gal and Laurence.(33) However, for the current study, our interest is only in the relative changes of the Li$^+$ chemical environment as a function of anion composition in a single aprotic solvent (both for DME and DMSO), such that $^7$Li NMR provides useful qualitative, if not quantitative, values for comparison.

In order to provide a quantitative basis for the role played by the electrolyte anion, we present a revised thermodynamic model for the solution electrochemical process. The solution mediated electrochemical growth of Li$_2$O$_2$ is triggered by the dissolution reaction given in Eq. (4). The free energy change involved in this dissolution reaction is given by:

$$\Delta G_{sol} = G_{Li^+_{sol}} + G_{O^-_{2,sol}} - G_{LiO^*_2} \quad (5)$$

where $G_{Li^+_{sol}}$ is the free energy of the Li$^+$ ions in the electrolyte, $G_{O^-_{2,sol}}$ is the free energy of $O_2^-$ ions in the electrolyte and $G_{LiO^*_2}$ is the free energy of the adsorbed LiO$_2$ on the Li$_2$O$_2$ surface during discharge.

To understand the role of salt anion on the equilibrium of the dissolution reaction, we need to explore the stabilization of the solvated intermediates in the presence of the anion. The presence of the anion can influence the free energy of Li$^+$ ions. The free energy of the Li$^+$ ions, to a first approximation, is largely dependent on the species that are present in its first solvation shell.(34, 35) As reported above in Section (III), we explicitly model an electrolyte that contains a mixture of LiNO$_3$ and LiTFSI such that the total Li$^+$ concentration is maintained at 1M. The concentration of $O_2^-$ ions in the solution is expected to be much lower than the Li$^+$ and the salt anion.(10) Hence, to a first approximation, the solvation shell of Li$^+$ will comprise only of solvent molecules and salt anions (NO$_3^-$ and TFSI$^-$). The exact composition of the solvation shell will depend on the energetics of the interactions of the Li$^+$ ion with the solvent and the anions. To determine the composition of the first solvation shell and in turn the free energy of stabilization, we develop a modified Ising model for the site occupancy in the solvation shell of Li$^+$.(36) The Ising model formalism, originally developed to describe magnetism, provides a systematic basis for treating the energetics of interaction between Li$^+$ and the solvent and salt anions.(37)

In this model, we develop a site occupancy variable to describe each of the solvation shell sites of Li⁺. The Hamiltonian that governs the solvation shell of Li⁺ is given by:

$$H = h_1 \sum_{i=1}^{N} n_i + h_2 \sum_{i=1}^{N} m_i + h_3 \sum_{i=1}^{N} l_i + J_{11} \sum_{\langle i,j \rangle} n_i n_j + J_{22} \sum_{\langle i,j \rangle} m_i m_j + J_{33} \sum_{\langle i,j \rangle} l_i l_j$$

$$J_{12} \sum_{\langle i,j \rangle} n_i m_j + J_{21} \sum_{\langle i,j \rangle} m_i n_j + J_{13} \sum_{\langle i,j \rangle} n_i l_j + J_{31} \sum_{\langle i,j \rangle} l_i n_j + J_{23} \sum_{\langle i,j \rangle} m_i l_j + J_{32} \sum_{\langle i,j \rangle} l_i m_j$$

____(6)

where i = 1 to N represents a site in the solvation shell of a Li⁺ ions. $\langle i, j \rangle$ represents the nearest neighbour pair in the solvation shell. The occupation variables 'n', 'm' and 'l' represent the occupancy of a site by the solvent, the $NO_3^-$ anions and the TFSI⁻ anions respectively. For any site 'i' occupied by the solvent, $n_i = 1$, $m_i = 0$ and $l_i = 0$ and similarly for other cases. Thus at any given site, we use the assumption that $n_i + m_i + l_i = 1$, i.e. each site is occupied by either solvent or a salt anion. In our model, $h_1$ represents the interaction energy between a Li⁺ ion and a solvent, $h_2$ represents the interaction energy between a $NO_3^-$ anion and Li⁺ and $h_3$ represents the interaction energy between a TFSI⁻ anion and Li⁺. The coupling constant $J_{11}$ represents the interaction between neighboring solvent molecules in the Li⁺ solvation shell. Likewise $J_{22}$ and $J_{33}$ represents the interaction between neighboring NO3⁻ and neighboring TFSI⁻ anions, respectively. The symmetry assumption is invoked which yields $J_{12} = J_{21}$, $J_{13} = J_{31}$, $J_{23} = J_{32}$. The cross-coupling terms, $J_{12}$, $J_{13}$ and $J_{23}$ represent interactions between neighboring $NO_3^-$ and solvent molecules, neighboring TFSI- and solvent molecules and neighboring TFSI$^-$ and NO$_3^-$ anions, respectively.

The exact model is not easily analytically tractable; however, we can invoke the mean field approximation, described in the SI. The mean field approximation is valid under the assumption that the Li$^+$ ions are uniformly distributed in solution and each site in the solvation shell experiences an averaged effect of other species present in the electrolyte. The coordination number z of the solvation shell is expected to be independent of species (anions or solvent) occupying the solvation shell. The mean field approximation replaces the nearest neighbor interaction ($n_i n_j$) by the average interaction ($n_i \langle n \rangle$), where assuming spatial invariance, the average occupation of species in the shell can be defined as $\langle n \rangle = \frac{1}{N}\sum_{i=1}^{N}\langle n_i \rangle$.

The interaction term h$_1$ is dependent on the donating tendency of the solvent molecule to the Li$^+$ ions in solution. The free energy of Li$^+$ ions can be expressed in terms of the half wave potential of Li/Li$^+$ couple and it has been shown that the half wave potential of Li/Li$^+$ couple is a function of the DN of the solvent.(38) Hence the Li$^+$-solvent interaction energetics (h$_1$) can be expressed as a function of the DN of the solvent. Similarly, we assume that the terms h$_2$ and h$_3$ can be expressed as a function of the DN. There is an additional contribution to h$_2$ and h$_3$ that depends on the concentration of the NO$_3^-$ and TFSI- anions. This arises due to a change in the reference chemical potential of the NO$_3^-$ and TFSI- anions to account for the configurational entropy associated with that concentration. The coupling constant J$_{11}$ is a weak attractive van der Waal interaction between solvent molecules, and is estimated to be an order of magnitude less than the donor interactions h$_1$, h$_2$ and h$_3$.

The constants J₂₂, J₃₃ and J₂₃ are representative of the repulsive interaction between neighboring anions in the Li⁺ solvation shell and are of the same order of magnitude as h₁, h₂. The coupling constants J₁₂, J₁₃ for the interaction between a solvent molecule and the respective anion can be described by the electron accepting tendency of the solvent and can therefore be determined by the solvent's AN. As we are accounting for the coupling constants in terms of the overall donating and accepting tendencies of the solvent, the overall coordination number is already included in the model, i.e., z = 1.

Solving Eqs. (S5-S7) described in the SI, we derive analytical expressions for the average occupation number of the solvent molecules and the anions in the first solvation shell of Li⁺ ion. These depend on the DN of the solvent and anion and also on the anion concentration. From the occupation numbers, we can determine the overall free energy of Li⁺ ions in solution using the mean-field relation:

$$G_{Li^+} = \langle n \rangle h_1(DN_{sol}) + \langle m \rangle h_2(DN_{NO_3^-}, c_{NO_3^-}) + \langle l \rangle h_3(DN_{TSFI^-}, c_{TFSI^-}) \quad\quad (9)$$

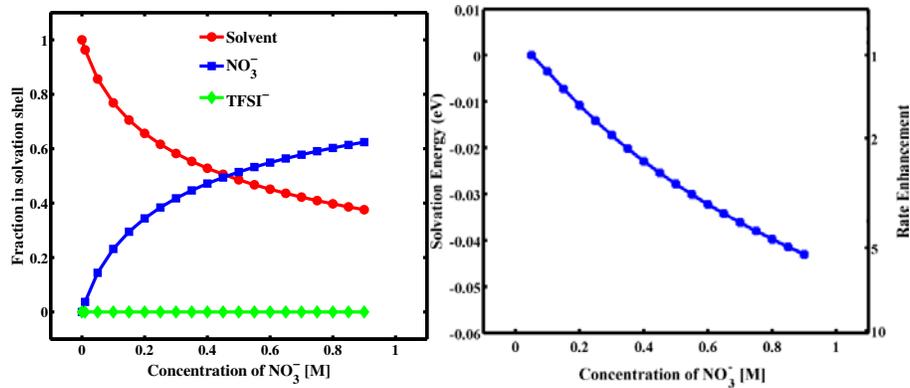

*Figure 4.* a) *The occupations of the solvent (red line), TFSI⁻ (green line) and $NO_3^-$ (blue line) in the Li⁺ solvation shell and* **b)** *the Li⁺ solvation energy (eV) as a function of the concentration of the $NO_3^-$ anion. The rate enhancement of the solution process, $r_S \sim exp\left(\frac{-\Delta G_{sol}}{kT}\right)$, is marked on the right y-axis of Figure 4b. The Li⁺ free energy is normalized relative to that of the case with 1M LiTFSI. There is a rate enhancement, $r_S$, by a factor of ~4 as the concentration of $NO_3^-$ is increased from 0.1M to 0.5M.*

The developed model requires the DN of $NO_3^-$ and TFSI- anions to determine the occupation numbers. We use the values reported by Schmeisser et al. using the indirect measurement of DN of anions using Na NMR spectroscopy on ionic liquids comprised of the anion and imidazolium cations of varying chain lengths.(39) Although an exact quantitative relationship and correspondence between these measurements to salt anions is still under debate, we believe the trends can be well captured from these values. Schmeisser et al. find that TFSI- has a very low DN of 11.2 while $NO_3^-$ has a DN of 22.2.(39) Using these values, we can determine the occupation shell of Li+ as a function of the $NO_3^-$ anion concentration and this shown in Figure 4a. The solvation shell is completely dominated by DME and $NO_3^-$ anion. As the $NO_3^-$ anion has a significantly higher DN than DME (DN=20), we observe a strong concentration dependence on $NO_3^-$ anion. This suggests that increasing $NO_3^-$ anion concentration will lead to a displacement of low DN solvents like DME in the Li+ solvation shell. As we increase the concentration of $NO_3^-$ in DME, higher number of $NO_3^-$ ions occupy the Li+ solvation shell until the electrostatic repulsion of $NO_3^-$ ions becomes dominant, leading to a saturation in the number of anions that occupy the first solvation shell.

The corresponding change in the free energy of Li+ as a function of $NO_3^-$ concentration is shown in Figure 4b. Due to a higher DN of the $NO_3^-$ anion, there is an overall increase in Li+ solvation energy; this is accompanied by an enhancement of the rate of the solution process given by $r_S \sim exp\left(\frac{-\Delta G_{sol}}{kT}\right)$. This shows that at 0.5 M LiNO$_3$/LiTFSI, we would expect a ~4 fold enhancement in the rate of the solution process.

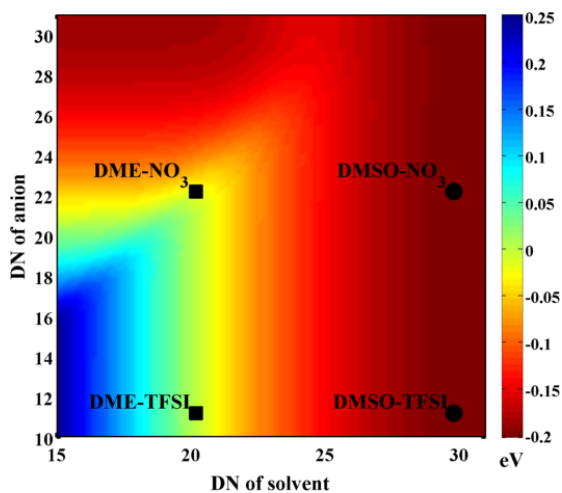

***Figure 5.*** *Contour plot showing the free energy of Li⁺ for electrolytes with varying DN of the solvent and salt anion. The free energy is normalized relative to that of DME and 1M LiTFSI. The electrolyte is considered to be a 50-50 mixture of LiTFSI and a salt consisting of Li⁺ and the salt anion in the corresponding solvent. The blue region corresponds to those electrolytes incapable of triggering the solution process while the red region corresponds to those that can trigger the solution process.*

The model developed can be generalized to map out the entire space of electrolyte design. A contour map of the Li$^+$ stabilization as a function of varying DN of solvent and anion is shown in Figure 5. This generalization analysis assumes a constant AN chosen to be the average of DME and DMSO and a 50:50 salt blend of LiTFSI and a varying electrolyte anion. The contour map shows that there is an enhancement in using low DN solvent, such as DME, and high DN salt anion, such as $NO_3^-$ ions. However, an interesting prediction of this generalized analysis is that there is no benefit in electrolyte blending using moderate DN salt anions in a high DN solvent such as DMSO. A more detailed analysis of the DMSO case is reported in the SI. This suggests that there is no ion-pair formation in a DMSO solvent and hence, almost no associated change in discharge capacity. This is in excellent agreement with the experiments presented in Figure 3.

The contour map provides a rational basis for selection of the total electrolyte, i.e. solvent and anion. An important conclusion of this contour map is there is minimal capacity enhancement by changing the electrolyte anion in high DN solvents. However, there is tremendous scope in tuning the electrolyte anion in low DN solvents to obtain high discharge capacities. This provides a route forward to avoid the unfavorable trade-off observed in high DN solvents where increased capacity comes at the expense of rechargeability, as is observed in DMSO.(17, 18, 40-42)

**IV. Conclusions**

In conclusion, we have demonstrated $Li^+$ counterion influence on promoting the solubility of electrochemical intermediates during a $Li-O_2$ battery discharge without further compromising electrolyte stability. Specifically, $Li-O_2$ batteries employing electrolytes of $LiNO_3$ and LiTFSI in DME display increasing capacity and increasing toroid formation with increasing $LiNO_3$ concentration. We ascribe intermediate solubility to enhanced stability of $Li^+$ in solution by anions with higher effective donor numbers than that of the solvent, thereby also allowing increased stability of the electrochemically formed anion, $O_2^-$, in solution. This strategy can be easily combined with current efforts to identify novel, stable electrolytes, including those in which organic molecules are entirely removed from the electrolyte (a fascinating idea being explored by researchers at Liox(43)), to develop a practical electrolyte that could enable high energy, long-life Li-air batteries. Further, we have developed a generalized model that predicts $Li^+$ solvation sphere occupation and the resulting stability of Li$^+$ in electrolytic solutions. We envision this strategy for intermediate stabilization to be generally applicable to numerous nonaqueous systems in which stabilization of desired intermediates may lead to improved electrochemical efficiency. For example, in Li-S batteries, polysulfide intermediate speciation could potentially be controlled by simply tuning the Li$^+$ anion, perhaps providing a route for increasing sulfur utilization.

**Acknowledgements**

The authors wish to thank Alan Luntz, Dan Addison, Hilda Buss, Jessica Nichols, and Christopher Dekmezian for helpful discussions and guidance on characterization of the materials studied here. The work at UC, Berkeley was supported in part by previous work performed through the Laboratory Directed Research and Development Program of Lawrence Berkeley National Laboratory under U.S. Department of Energy Contract No. DE-AC02-05CH11231, and by start-up funds through the Department of Chemical and Biomolecular Engineering at UC, Berkeley. The work at Carnegie Mellon University is supported through start-up funds through the Department of Mechanical Engineering at Carnegie Mellon University.

**V. References**

1. Bruce PG, Freunberger SA, Hardwick LJ, & Tarascon J-M (2012) Li-O$_2$ and Li-S batteries with high energy storage. *Nature Mater.* 11(1):19-29.
2. Christensen J, *et al.* (2012) A critical review of Li/air batteries. *J. Electrochem. Soc.* 159(2):R1-R30.
3. Luntz AC & McCloskey BD (2014) Nonaqueous Li–air batteries: a status report. *Chemical Reviews* 114(23):11721-11750.
4. Lu J, *et al.* (2014) Aprotic and aqueous Li–O$_2$ batteries. *Chemical Reviews* 114(11):5611-5640.

5. Viswanathan V, *et al.* (2011) Electrical conductivity in $Li_2O_2$ and its role in determining capacity limitations in non-aqueous $Li-O_2$ batteries. *J. Chem. Phys.* 135(21):214704.
6. Albertus P, *et al.* (2011) Identifying capacity limitations in the Li/oxygen battery using experiments and modeling. *J. Electrochem. Soc.* 158(3):A343-A351.
7. Meini S, Piana M, Beyer H, Schwämmlein J, & Gasteiger HA (2012) Effect of carbon surface area on first discharge capacity of $Li-O_2$ cathodes and cycle-life behavior in ether-based electrolytes. *J. Electrochem. Soc.* 159(12):A2135-A2142.
8. Gerbig O, Merkle R, & Maier J (2013) Electron and ion transport in $Li_2O_2$. *Advanced Materials* 25(22):3129-3133.
9. Højberg J, *et al.* (2015) An electrochemical impedance spectroscopy investigation of the overpotentials in Li–$O_2$ batteries. *ACS Appl. Mater. Interfaces* 7(7):4039-4047.
10. Aetukuri NB, *et al.* (2015) Solvating additives drive solution-mediated electrochemistry and enhance toroid growth in non-aqueous Li–$O_2$ batteries. *Nature Chem.* 7(1):50-56.
11. Meini S, Piana M, Tsiouvaras N, Garsuch A, & Gasteiger HA (2012) The effect of water on the discharge capacity of a non-catalyzed carbon cathode for Li-$O_2$ batteries. *Electrochem. Solid-State Lett.* 15(4):A45-A48.
12. Schwenke KU, Metzger M, Restle T, Piana M, & Gasteiger HA (2015) The influence of water and protons on $Li_2O_2$ crystal growth in aprotic Li-$O_2$ cells. *J. Electrochem. Soc.* 162(4):A573-A584.
13. Mayer U, Gutmann V, & Gerger W (1975) The acceptor number — A quantitative empirical parameter for the electrophilic properties of solvents. *Monatshefte für Chemie* 106(6):1235-1257.
14. Gutmann V & Wychera E (1966) Coordination reactions in non aqueous solutions - The role of the donor strength. *Inorganic and Nuclear Chemistry Letters* 2(9):257-260.
15. Johnson L, *et al.* (2014) The role of $LiO_2$ solubility in $O_2$ reduction in aprotic solvents and its consequences for Li–$O_2$ batteries. *Nature Chem.* 6(12):1091-1099.
16. Cho MH, *et al.* (2014) The effects of moisture contamination in the Li-$O_2$ battery. *J. Power Sources* 268:565-574.
17. McCloskey BD, *et al.* (2012) Limitations in rechargeability of Li-$O_2$ batteries and possible origins. *J. Phys. Chem. Lett.*:3043-3047.
18. Khetan A, Luntz A, & Viswanathan V (2015) Trade-offs in capacity and rechargeability in nonaqueous Li–$O_2$ batteries: solution-driven growth versus nucleophilic stability. *J. Phys. Chem. Lett.*:1254-1259.
19. Walker W, *et al.* (2013) A rechargeable Li–$O_2$ battery using a lithium nitrate/N,N-dimethylacetamide electrolyte. *J. Am. Chem. Soc.* 135(6):2076-2079.
20. Uddin J, *et al.* (2013) Lithium nitrate as regenerable SEI stabilizing agent for rechargeable Li/$O_2$ batteries. *J. Phys. Chem. Lett.* 4(21):3760-3765.

21. McCloskey BD, Bethune DS, Shelby RM, Girishkumar G, & Luntz AC (2011) Solvents' critical role in nonaqueous lithium–oxygen battery electrochemistry. *J. Phys. Chem. Lett.* 2(10):1161-1166.
22. McCloskey BD, Scheffler R, Speidel A, Girishkumar G, & Luntz AC (2012) On the mechanism of nonaqueous Li–$O_2$ electrochemistry on C and its kinetic overpotentials: some implications for Li–air batteries. *J. Phys. Chem. C* 116(45):23897-23905.
23. Laoire CO, Mukerjee S, Abraham KM, Plichta EJ, & Hendrickson MA (2010) Influence of nonaqueous solvents on the electrochemistry of oxygen in the rechargeable lithium-air battery. *J. Phys. Chem. C* 114(19):9178-9186.
24. Laoire CO, Mukerjee S, Abraham KM, Plichta EJ, & Hendrickson MA (2009) Elucidating the mechanism of oxygen reduction for lithium-air battery applications. *J. Phys. Chem. C* 113(46):20127-20134.
25. Peng Z, *et al.* (2011) Oxygen reactions in a non-aqueous $Li^+$ electrolyte. *Angew. Chem., Int. Ed.* 50(28):6351-6355.
26. Lu YC, *et al.* (2013) Lithium-oxygen batteries: bridging mechanistic understanding and battery performance. *Energy & Environmental Science* 6(3):750-768.
27. McCloskey BD, *et al.* (2013) Combining accurate $O_2$ and $Li_2O_2$ assays to separate discharge and charge stability limitations in nonaqueous Li-$O_2$ batteries. *Journal of Physical Chemistry Letters* 4:2989- 2993.
28. Adams BD, *et al.* (2013) Current density dependence of peroxide formation in the Li-$O_2$ battery and its effect on charge. *Energy & Environmental Science* 6(6):1772-1778.
29. Erlich RH & Popov AI (1971) Spectroscopic studies of ionic solvation. X. Study of the solvation of sodium ions in nonaqueous solvents by sodium-23 nuclear magnetic resonance. *J. Am. Chem. Soc.* 93(22):5620-5623.
30. Cretzmeyer JW (1963) Process for producing anhydrous lithium perchlorate, Patent #US3075827.
31. Cahen YM, Handy PR, Roach ET, & Popov AI (1975) Spectroscopic studies of ionic solvation. XVI. Lithium-7 and chlorine-35 nuclear magnetic resonance studies in various solvents. *J. Phys. Chem.* 79(1):80-85.
32. Linert W, Jameson RF, & Taha A (1993) Donor numbers of anions in solution: the use of solvatochromic Lewis acid-base indicators. *Journal of the Chemical Society, Dalton Transactions* (21):3181-3186.
33. Gal J-F & Laurence C (2013) Comment on the article "Gutmann donor and acceptor numbers for ionic liquids" by M. Schmeisser, P. Illner, R. Puchta, A. Zahl, and R. van Eldik (Chem. Eur. J. 2012, 18, 10969–10982). *Chemistry – A European Journal* 19(49):16832-16834.
34. Marcus Y (1991) Thermodynamics of solvation of ions. Part 5.-Gibbs free energy of hydration at 298.15 K. *J. Chem. Soc., Faraday Trans.* 87(18):2995-2999.
35. Bryantsev VS, Diallo MS, & Goddard Iii WA (2008) Calculation of solvation free energies of charged solutes using mixed cluster/continuum models. *J. Phys. Chem. B* 112(32):9709-9719.

36. Sethna J (2006) *Statistical mechanics: entropy, order parameters, and complexity* (Oxford University Press).
37. Ising E (1925) Beitrag zur Theorie des Ferromagnetismus. *Zeitschrift fur Physik* 31(1):253-258.
38. Gritzner G (1986) Solvent effects on half-wave potentials. *J. Phys. Chem.* 90(21):5478-5485.
39. Schmeisser M, Illner P, Puchta R, Zahl A, & van Eldik R (2012) Gutmann donor and acceptor numbers for ionic liquids. *Chemistry – A European Journal* 18(35):10969-10982.
40. Kwabi DG, *et al.* (2014) Chemical instability of dimethyl sulfoxide in lithium–air batteries. *J. Phys. Chem. Lett.* 5:2850-2856.
41. Sharon D, *et al.* (2013) Oxidation of dimethyl sulfoxide solutions by electrochemical reduction of oxygen. *J. Phys. Chem. Lett.* 4(18):3115-3119.
42. Younesi R, Norby P, & Vegge T (2014) A new look at the stability of dimethyl sulfoxide and acetonitrile in Li-O2 batteries. *ECS Electrochemistry Letters* 3(3):A15-A18.
43. Uddin J, Addison DD, Giordani V, Chase GV, & Walker W (2014) Alkali metal/oxygen batteries employing molten nitrate electrolytes. (Patent# WO 2014153551 A1).

# Enhancing electrochemical intermediate solvation through electrolyte anion selection to increase nonaqueous Li-O$_2$ battery capacity

Colin M. Burke[1,2], Vikram Pande[3], Abhishek Khetan[4], Venkatasubramanian Viswanathan[3]* and Bryan D. McCloskey[1,2]*

[1]*Department of Chemical and Biomolecular Engineering, University of California, Berkeley, CA, 94720*
[2]*Environmental Energy Technologies Division, Lawrence Berkeley National Laboratory, Berkeley, CA, 94720*
[3]*Department of Mechanical Engineering, Carnegie Mellon University, Pittsburgh, PA, 15213*
[4]*Institute for Combustion Technology, RWTH, Aachen 52056, Germany*

*bmcclosk@berkeley.edu, venkvis@cmu.edu

## Supporting Information

**Materials**

Lithium nitrate was purchased from Sigma Aldrich and was dried under vacuum in a heated glove box antechamber at 150°C for 24 hours before use. Lithium bis(trifluoromethane) sulfonimide (LiTFSI), 1,2-dimethoxyethane, and dimethyl sulfoxide were purchased from BASF and used as received. Whatman QM-A glass fiber filters were purchased from VWR. PTFE (60 wt% dispersion in H$_2$O) was purchased from Sigma Aldrich. Vulcan XC72 was purchased from the Fuel Cell Store and was filtered through a 60 mesh screen. T316 stainless steel 120 mesh, with wire diameter 0.0026", was purchased from TWP Inc. Research-grade oxygen and argon were purchased from Praxair. 99% $^{18}O_2$ was purchased from Sigma Aldrich. All electrolyte, cell, and NMR sample preparation was completed in an argon-filled glove box with <0.1ppm O$_2$ and <0.1ppm H$_2$O.

In preparing the electrolytes, we found the solubility limit of LiNO$_3$ in DME is approximately 1M. The capacity variability of the 0.7M LiNO$_3$ cell in Figure 1 is likely due to concentration polarization effects in the electrolyte, as 0.7M begins to approach the solubility limit of LiNO$_3$ in DME, such that LiNO$_3$ precipitation at the anode may occur. The LiNO$_3$ concentration range (<0.7M) we report was limited by this effect.

**Cathode preparation**

Cathodes were prepared via a similar method to that described previously. (1) A mixture of 3:1 w:w ratio of Vulcan XC72 to PTFE binder in IPA and water (4:1 water:IPA; and 15mL total for 400mg C) was sonicated for 30 seconds and homogenized for 6 minutes. A Badger model 250 air-sprayer was used to spray the slurry onto a piece of stainless steel mesh roughly 4.5" square, which had been rinsed with IPA and acetone and dried at 150°C for ten minutes prior to use. After letting the carbon air-dry, 12mm diameter cathodes were punched out, rinsed with IPA and acetone, and dried at 150°C under vacuum for at least 12 hours. The cathodes were then transferred, while hot, into the glove box, and stored on a hot plate at 200°C.

**Carbon loading**

Cell capacities depend on carbon loading. To keep carbon loading consistent, the cathodes used for any particular data set, such as the capacity measurements displayed in Figure 1 or the cathode morphology images in Figure 2, were all from the same batch of spray-coated cathodes. As a control, the capacity measurements presented in Figure 1 were repeated with a second batch of cathodes. All capacities changed proportionally (a slight increase for all cases), with the capacity of the cell using 0.5M $LiNO_3$/0.5M LiTFSI in DME maintaining just over a three-fold increase from the capacity of the cell employing 1M LiTFSI in DME. Cathodes contained on average 1.5-2 mg/cm² carbon.

**Cell preparation**

The Li-$O_2$ cells used throughout this paper followed the same Swagelok design as described previously.(2) All cells employed a 7/16" diameter lithium foil, a 1/2" diameter Whatman QM-A glass fiber separator, a 12mm diameter cathode of Vulcan XC72 carbon on stainless steel mesh, and a 1mm thick, 1/2" diameter stainless steel ring. The QM-A separators, like the cathodes, were rinsed with IPA and acetone and dried under vacuum at 150°C under vacuum for at least 12 hours before being transferred to the glove box and stored on a hot plate at 200°C. Each battery contained 80 μL of electrolyte.

**Scanning electron microscopy**

Discharged cathodes were characterized via scanning electron microscopy immediately after discharge. After replacing discharged cells' headspaces with argon, the cells were transferred into the glove box, and the cathodes were removed. The cathodes were each rinsed with two 1 mL aliquots of DME and were subsequently dried under vacuum for at least five minutes in the glove box antechamber. The cathodes were then sealed in septa vials, removed from the glove box, and taken to the SEM. Immediately before imaging, the cathodes were removed from the argon-filled septa vials, placed on carbon tape on the SEM holder, and inserted into the SEM. From discharge completion to SEM insertion was typically one hour. From removing the cathodes from the septa vials to SEM insertion was typically less than 30 seconds. SEM was performed on a JEOL JSM-7500f.

**Titration**

The $Li_2O_2$ titration protocol used here followed that described previously.(1) After replacing the discharged cells' headspaces with argon, the cells were transferred into the glove box, and the cathodes were removed. Each cathode was placed in a 20mL septa sealed vial. The cathodes were placed under vacuum for at least three minutes to evaporate any residual solvent before the vial caps were tightly sealed and the vials were removed from the glove box. 2 mL of ultrapure water (18.2 MΩ cm, Millipore) was injected through the septa to react the $Li_2O_2$ into LiOH and $H_2O_2$. LiOH was quantified by using phenolphthalein as an indicator and HCl as a titrant. $H_2O_2$ was quantified via an iodometric titration employing potassium iodide, sulfuric acid, and a molybdate catalyst solution to create $I_2$ and sodium thiosulfate as a titrant for the $I_2$. The $Li_2O_2$ percent yield is defined as the amount of $Li_2O_2$ formed during discharge, as quantified via an iodometric titration, to the amount of $Li_2O_2$ expected from coulometry, assuming an ideal $2e^-/Li_2O_2$ process.(1)  Of note, no titratable $I_2$ was observed from titrations on the separator alone, indicating that any $NO_2^-$ formed at Li metal anode does not interfere with the peroxide titration accuracy.

**Nuclear Magnetic Resonance Spectroscopy**

$^7$Li and $^{23}$Na nuclear magnetic resonance spectroscopy measurements were completed on a Bruker AM-400 magnet with a 5mm Z-gradient broad brand probe. Reference samples, those employing a chloride salt in $D_2O$, were prepared outside the glove box and were flame-sealed in melting point capillaries. Electrolyte samples were prepared inside the glove box and were placed, along with a reference capillary, in a Wilmad screw-cap NMR tube. All reference samples were 3M of the chloride salt in $D_2O$. For the $^{23}$Na NMR, 0.2M $NaClO_4$ was added. These molarities were taken from Schmeisser et al.(3) $^{23}$Na NMR spectra are not reported here because of the poor solubility of $NaNO_3$ in DME. NMR spectra of only the reference and only the sample were taken to verify the identity of each peak. A representative $^7$Li NMR spectrum of a 0.5M LiTFSI/0.5M $LiNO_3$ in DME electrolyte is shown in Figure S7.

**Water controls**

With the solubility mechanism confirmed, it was important to check that the increased solubility was indeed due to DN effects of $NO_3^-$ and not another experimental artifact, in particular water contamination. As a first control, the key experiments were repeated with new electrolyte solutions that used lithium nitrate powder that had been dried a second time under vacuum in a heated glove box antechamber at 150°C for 24 hours. All repeated experiments gave consistent results with their original counterparts. As a second control, water levels in the electrolytes were measured via Karl Fischer titration. All electrolytes had less than 70ppm water, although the 1M LiTFSI (0M $LiNO_3$) had <10ppm water content. Therefore, a battery employing an electrolyte of 1M LiTFSI in DME with 70ppm of water was discharged at 450 μA/cm$^2$ to 2V, analogous to the batteries in Figure 1. This battery saw an increase in capacity of 30% compared to a battery with nominally anhydrous 1M LiTFSI in DME, much smaller than the four-fold increase displayed in Figure 1 from the cell employing the 0.3M LiTFSI/0.7M $LiNO_3$ electrolyte.

**Ising Model for Li⁺ solvation shell**

The composition of the first solvation shell is determined using a modified Ising model. The energetic interactions in the solvation shell of Li⁺ can be described by a model Hamiltonian, H, given by:

$$H = h_1 \sum_{i=1}^{N} n_i + h_2 \sum_{i=1}^{N} m_i + h_3 \sum_{i=1}^{N} l_i + J_{11} \sum_{\langle i,j \rangle} n_i n_j + J_{22} \sum_{\langle i,j \rangle} m_i m_j + J_{33} \sum_{\langle i,j \rangle} l_i l_j$$

$$J_{12} \sum_{\langle i,j \rangle} n_i m_j + J_{21} \sum_{\langle i,j \rangle} m_i n_j + J_{13} \sum_{\langle i,j \rangle} n_i l_j + J_{31} \sum_{\langle i,j \rangle} l_i n_j + J_{23} \sum_{\langle i,j \rangle} m_i l_j + J_{32} \sum_{\langle i,j \rangle} l_i m_j$$

____(S1)

An analytical solution to the exact Hamiltonian, H, is difficult. We invoke the mean field approximation which replaces the nearest neighbor interaction ($n_i n_j$) by the average interaction ($n_i \langle n \rangle$). The mean-field Hamiltonian, H is given by

$$H = h_1 \sum_{i=1}^{N} n_i + h_2 \sum_{i=1}^{N} m_i + h_3 \sum_{i=1}^{N} l_i + J_{11} z \langle n \rangle \sum_{i=1}^{N} n_i + J_{22} z \langle m \rangle \sum_{i=1}^{N} m_i + J_{33} z \langle l \rangle \sum_{i=1}^{N} l_i$$

$$+ J_{12} \frac{z}{2} \langle m \rangle \sum_{i=1}^{N} n_i + J_{21} \frac{z}{2} \langle n \rangle \sum_{i=1}^{N} m_i + J_{13} \frac{z}{2} \langle l \rangle \sum_{i=1}^{N} n_i + J_{31} \frac{z}{2} \langle n \rangle \sum_{i=1}^{N} l_i$$

$$+ J_{23} \frac{z}{2} \langle l \rangle \sum_{i=1}^{N} m_i + J_{32} \frac{z}{2} \langle m \rangle \sum_{i=1}^{N} l_i \quad \_\_\_\_(S2)$$

The average occupations at a site can be found by performing an average on each occupation variable using the mean field energy in the Boltzmann weights as probabilities.

$$\langle n \rangle = \frac{\sum_{n=0,1} \sum_{m=0,1} \sum_{l=0,1} n \exp\left(\frac{-H}{kT}\right)}{\sum_{n=0,1} \sum_{m=0,1} \sum_{l=0,1} \exp\left(\frac{-H}{kT}\right)} \quad \_\_\_\_(S3.a)$$

$$\langle m \rangle = \frac{\sum_{n=0,1} \sum_{m=0,1} \sum_{l=0,1} m \exp\left(\frac{-H}{kT}\right)}{\sum_{n=0,1} \sum_{m=0,1} \sum_{l=0,1} \exp\left(\frac{-H}{kT}\right)} \quad \_\_\_\_(S3.b)$$

$$\langle l \rangle = \frac{\sum_{n=0,1}\sum_{m=0,1}\sum_{l=0,1} l \exp\left(\frac{-H}{kT}\right)}{\sum_{n=0,1}\sum_{m=0,1}\sum_{l=0,1} \exp\left(\frac{-H}{kT}\right)} \quad\_\_\_(S3.c)$$

From these expressions, we get the average occupation of solvent and salt anion at a site in the Li$^+$ solvation shell as given below:

$$\langle n \rangle = \frac{\exp\left(\frac{\left(-h_1-J_{11}z\langle n\rangle - J_{12}\frac{z}{2}\langle m\rangle - J_{13}\frac{z}{2}\langle l\rangle\right)}{kT}\right)}{\exp\left(\frac{\left(-h_1-J_{11}z\langle n\rangle - J_{12}\frac{z}{2}\langle m\rangle - J_{13}\frac{z}{2}\langle l\rangle\right)}{kT}\right) + \exp\left(\frac{\left(-h_2-J_{22}z\langle m\rangle - J_{21}\frac{z}{2}\langle n\rangle - J_{23}\frac{z}{2}\langle l\rangle\right)}{kT}\right) + \exp\left(\frac{\left(-h_3-J_{33}z\langle l\rangle - J_{31}\frac{z}{2}\langle n\rangle - J_{32}\frac{z}{2}\langle m\rangle\right)}{kT}\right)} \quad\_\_\_(S4.a)$$

$$\langle m \rangle = \frac{\exp\left(\frac{\left(-h_2-J_{22}z\langle m\rangle - J_{21}\frac{z}{2}\langle n\rangle - J_{23}\frac{z}{2}\langle l\rangle\right)}{kT}\right)}{\exp\left(\frac{\left(-h_1-J_{11}z\langle n\rangle - J_{12}\frac{z}{2}\langle m\rangle - J_{13}\frac{z}{2}\langle l\rangle\right)}{kT}\right) + \exp\left(\frac{\left(-h_2-J_{22}z\langle m\rangle - J_{21}\frac{z}{2}\langle n\rangle - J_{23}\frac{z}{2}\langle l\rangle\right)}{kT}\right) + \exp\left(\frac{\left(-h_3-J_{33}z\langle l\rangle - J_{31}\frac{z}{2}\langle n\rangle - J_{32}\frac{z}{2}\langle m\rangle\right)}{kT}\right)} \quad\_\_\_(S4.b)$$

$$\langle l \rangle = \frac{\exp\left(\frac{\left(-h_3-J_{33}z\langle l\rangle - J_{31}\frac{z}{2}\langle n\rangle - J_{32}\frac{z}{2}\langle m\rangle\right)}{kT}\right)}{\exp\left(\frac{\left(-h_1-J_{11}z\langle n\rangle - J_{12}\frac{z}{2}\langle m\rangle - J_{13}\frac{z}{2}\langle l\rangle\right)}{kT}\right) + \exp\left(\frac{\left(-h_2-J_{22}z\langle m\rangle - J_{21}\frac{z}{2}\langle n\rangle - J_{23}\frac{z}{2}\langle l\rangle\right)}{kT}\right) + \exp\left(\frac{\left(-h_3-J_{33}z\langle l\rangle - J_{31}\frac{z}{2}\langle n\rangle - J_{32}\frac{z}{2}\langle m\rangle\right)}{kT}\right)} \quad\_\_\_(S4.c)$$

The interaction term $h_1$ is dependent on the donating tendency of the solvent molecule to the Li$^+$ ions in solution. This can be determined using the equilibrium between solvated Li$^+$ ions and metallic Li in a given solvent (HA), given by

$$Li_{(s)} \rightleftharpoons Li^+_{HA} + e^- \quad\_\_\_(S5)$$

The free energy of Li$^+$ ions can be expressed in terms of the half wave potential of the Li/Li$^+$ couple in that solvent as:

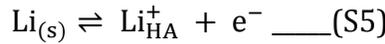

$$G_{Li^+_{HA}} = G_{Li_{(s)}} + eU_{Li/Li^+} \quad\_\_\_(S6)$$

It has been shown that the half wave potential of the Li/Li$^+$ couple is a function of the DN of the solvent. Similarly the interaction terms of the anions with Li$^+$ ($h_2$ and $h_3$) can be expressed as a function of their respective DNs and the solvent-anion interactions ($J_{12}$, $J_{21}$, $J_{13}$, $J_{31}$) can be expressed as a function of the AN of the solvent. The solvent-solvent interaction term ($J_{11}$) represents weak van der Waal's forces and is hence is chosen to be -0.01 eV. The strong anion repulsion terms ($J_{22}$, $J_{23}$, $J_{32}$, $J_{33}$) are chosen to be 0.1 eV.

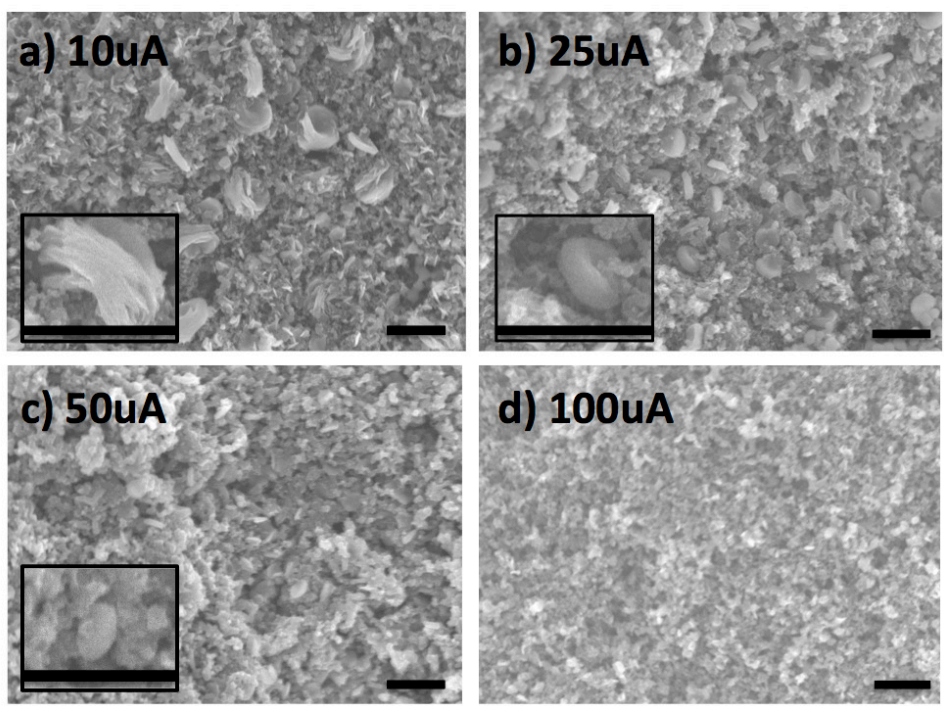

*Figure S1.* Scanning electron microscope images of carbon cathodes from cells employing 0.5M LiNO$_3$/0.5M LiTFSI in DME discharged to 1mAh capacity at currents of 10μA **(a)**, 25μA **(b)**, 50μA **(c)**, and 100μA **(d)**. Total electrode area was 1.1 cm$^2$ (12 mm diameter).

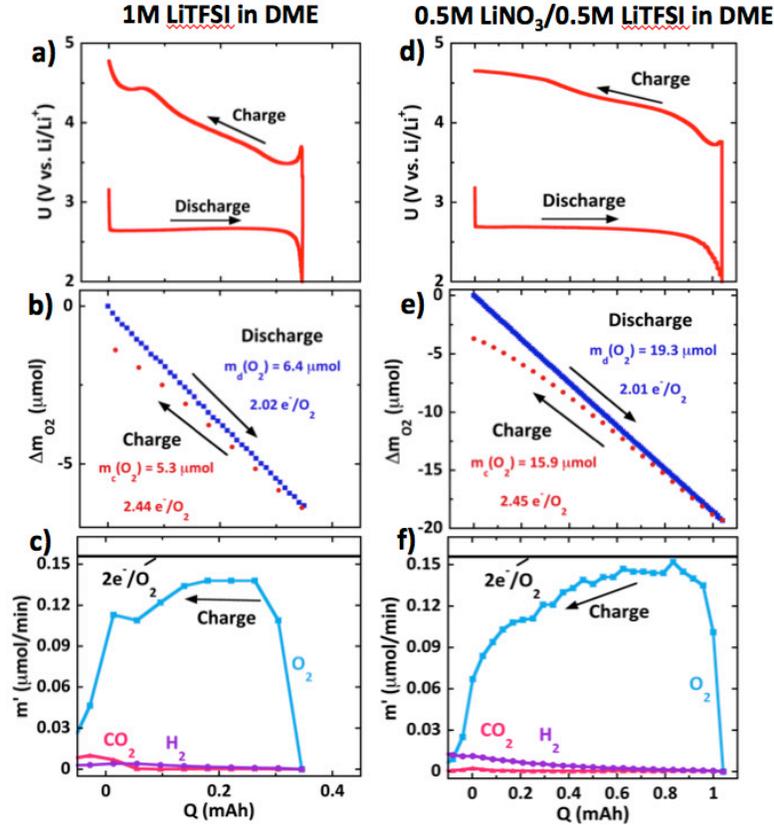

***Figure S2. (a, d)*** *Galvanostatic discharge-charge curves for cells employing **(a-c)** 1M LiTFSI in DME and **(d-f)** 0.5M LiNO₃/0.5M LiTFSI in DME. **(b, d)** Oxygen consumption during discharge and evolution during charge. **(c, f)** Gas evolution rates for $H_2$, $CO_2$, and $O_2$ during cell charge. These were the only gases found to evolve during charge.*

Figures S2(a) and S2(d) present galvanostatic discharge-charge profiles for cells employing 1M LiTFSI and 0.5M LiNO₃/0.5M LiTFSI. Discharge, or ORR, overpotentials are similar between the two electrolytes. The cell employing LiNO₃ salt, however, exhibited a higher initial charge overpotential, which likely arises from an electronic transport limitation to toroid decomposition, although further studies are necessary to confirm this hypothesis. Figure S6 shows that for cells charged from the same discharge capacity, this initial charge overpotential does increase with increased LiNO₃ concentration.

Figures S2(b) and S2(e) show oxygen consumption as a function of capacity during discharge and charge for these two cells. On both discharge and charge, both cells show nearly identical $e^-/O_2$, indicating nearly equivalent degrees of reversibility. Figure S2(c) and S2(f) display gas evolution on charge from DEMS.

Oxygen was clearly the dominant gas evolved for both cells, and in nearly identical quantities, with oxygen was 92% of the gases evolved from the 0M LiNO$_3$ (1M LiTFSI), and 93% of the gases from the 0.5M LiNO$_3$ (0.5M LiTFSI) cell. Small quantities of carbon dioxide and hydrogen were evolved from each cell toward the end of charging, with the 0M LiNO$_3$ evolving slightly more carbon dioxide. The 0.5M LiNO$_3$ cell evolved slightly more hydrogen.

**Model Results**

The occupation variables derived the model for an electrolyte using DMSO as a solvent is shown in Figure S3. This shows that in DMSO, which is a high DN solvent, $NO_3^-$ anion is unable to replace DMSO from the solvation shell. The accompanied change in free energy of Li$^+$ as a function of $NO_3^-$ anion concentration is shown in Figure S4. This shows that there is almost no change in discharge capacity consistent with the experiments reported here.

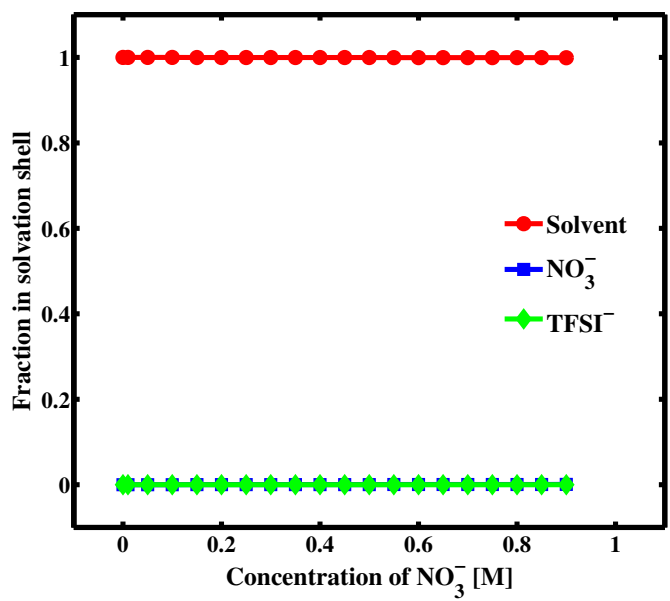

*Figure S3.* *The change in occupations of the solvent DMSO (red line), TFSI$^-$ (green line) and $NO_3^-$ (blue line) in the Li$^+$ solvation shell as we change the concentration of the $NO_3^-$ anion. TFSI$^-$ and $NO_3^-$, both have lower DNs (11.2 and 22.2) compared to the solvent DMSO (29.8). Hence they cannot replace DMSO in the solvation shell. Hence even with increasing $NO_3^-$ concentration, DMSO completely occupies the solvation of Li$^+$.*

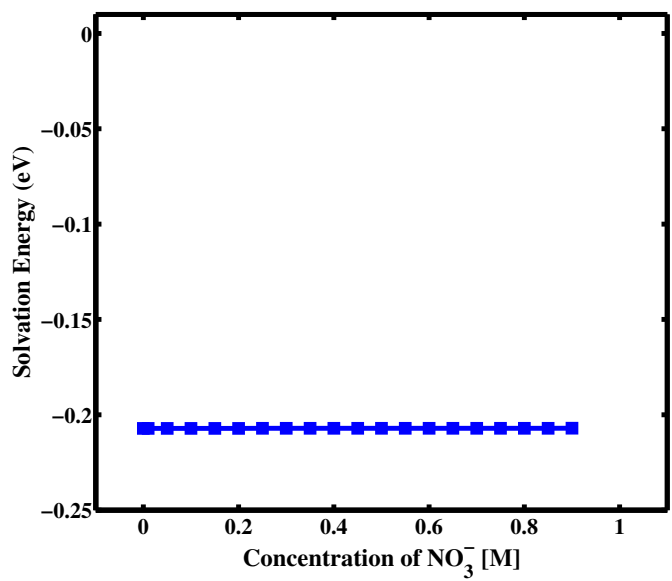

**Figure S4.** The Li$^+$ solvation energy (eV) as a function of the concentration of the NO$_3^-$ anion. The solvent used is DMSO and the salt is mixture with different concentrations of LiNO$_3$ and LiTFSI such that the Li$^+$ concentration is maintained at 1M. The Li$^+$ free energy of Li$^+$, calculated relative to DME and 1M LiTFSI, is independent of the NO$_3^-$ concentration when the solvent is DMSO. As a result, the solution rate enhancement is solely due to the high DN solvent DMSO.

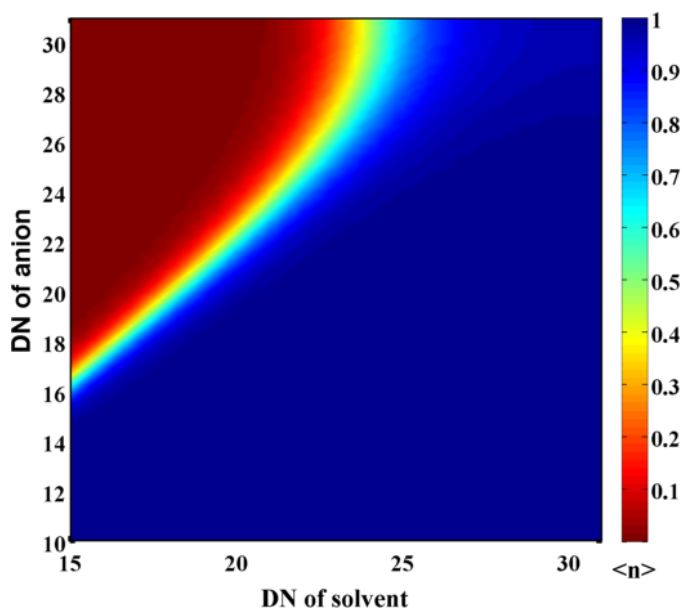

**Figure S5.** Contour plot showing the occupation in Li⁺ solvation shell of the solvent for varying DN of the solvent and salt anion. The salt anions shown are such that they have DN between 10 and 30 while the solvents shown have DN between 15-30. The electrolyte is considered to be a 50-50 mixture of LiTFSI and a salt consisting of Li⁺ and the salt anion in the corresponding solvent. The blue region represents high solvent occupation in the Li⁺ solvation shell while the red region shows high anion occupation.

**Generalized Electrolyte Design**

A generalized electrolyte design model is developed based on the Ising model discussed above. A contour map of the occupation of the solvent as a function of varying DN of solvent and anion is shown in Figure S5. This generalization analysis assumes a constant AN chosen to be the average of DME and DMSO and a 50:50 salt blend of LiTFSI and a varying electrolyte anion. The contour map shows that in a low DN solvent, utilizing a high DN anion leads to replacement of the solvent by the anion in the solvation shell. For a DN of 20.2 which corresponds to DME, an anion DN of ~23 leads to an equal amount of solvent and anion in the solvation shell. It is worth highlighting that the occupation is a stronger function of the DN of the solvent than that of the anion. For high DN (>25) solvents, the solvation shell is predominatly occupied by the solvent.

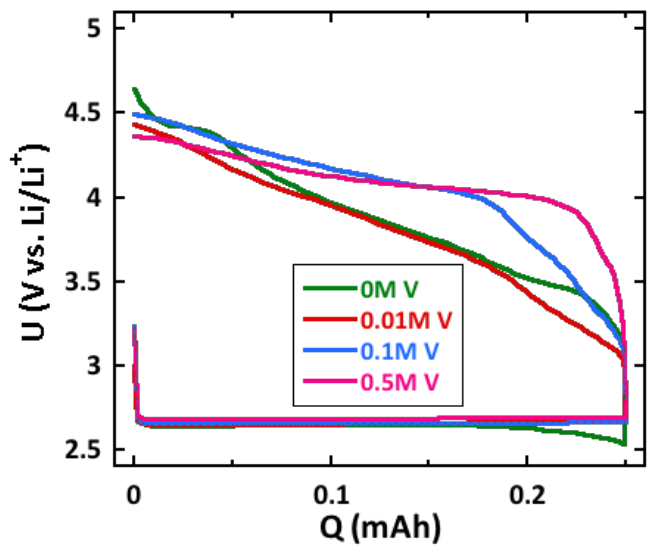

*Figure S6.* Representative discharge-charge profiles of cells of various LiNO$_3$ concentrations. . All cells were discharged at 450 μA/cm² to 0.23 mAh/cm². A 1.0M Li$^+$ concentration was used for all cell electrolytes, while the LiTFSI/LiNO$_3$ ratio was varied. The LiNO$_3$ concentration for each cell is provided in the figure legend.

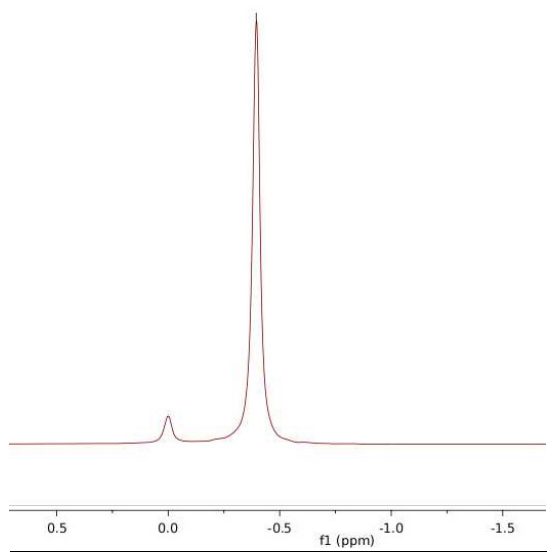

***Figure S7.*** *Representative $^7$Li NMR spectrum on 0.5M LiTFSI/0.5M LiNO$_3$ in DME. The chemical shift at 0 ppm corresponds to the Li shift of LiCl in D$_2$O, whereas the chemical shift at -0.39 ppm corresponds to the Li shift in the electrolyte.*

**References**

1. McCloskey BD*, et al.* (2013) Combining accurate O$_2$ and Li$_2$O$_2$ assays to separate discharge and charge stability limitations in nonaqueous Li-O$_2$ batteries. *Journal of Physical Chemistry Letters* 4:2989- 2993.
2. McCloskey BD, Bethune DS, Shelby RM, Girishkumar G, & Luntz AC (2011) Solvents' critical role in nonaqueous lithium–oxygen battery electrochemistry. *J. Phys. Chem. Lett.* 2(10):1161-1166.
3. Schmeisser M, Illner P, Puchta R, Zahl A, & van Eldik R (2012) Gutmann donor and acceptor numbers for ionic liquids. *Chemistry – A European Journal* 18(35):10969-10982.